\documentclass[preprint,showkeys,aps,draft]{revtex4}
\usepackage{epsf}
\usepackage{dcolumn}% Align table columns on decimal point
\usepackage{bm}% bold math
\everymath{\displaystyle}
\begin{document}
%%%%%%%%%%%%%%%%%%%%%%%%%%%%%%%%%%%%%%%%%%%%%%%%%%%%%%%%%%%%%%%%%%%%
% End of preamble and beginning of text.
%\pagestyle{empty}
%%%%%%%%%%%%%%%%%%%%%%%%%%%%%%%%%%%%%%%%%%%%%%%%%%%%%%%%%%%%%

\title{Foldy-Wouthuysen Transformation and Semiclassical Transition for
Relativistic Quantum Mechanics}

\author{Alexander J. Silenko}
%\email{silenko@inp.minsk.by}
\affiliation{Research Institute for Nuclear Problems, Belarusian
State University, Minsk 220030, Belarus}

\begin {abstract}
It is shown that the Foldy-Wouthuysen transformation for
relativistic particles in strong external fields provides the
possibility of obtaining a meaningful classical limit of the
relativistic quantum mechanics. The full agreement between quantum
and classical theories is proved. The coincidence of the
semiclassical equations of motion of particles and their spins
with the corresponding classical equations is established. The
Niels Bohr's correspondence principle is valid not only in the
limit of large spin quantum numbers but also for particles with
any spin as well as for spinless particles.
\end{abstract}
\keywords{Foldy-Wouthuysen transformation; relativistic quantum
mechanics; semiclassical transition; correspondence principle}
\maketitle

\section{Introduction}

The Foldy-Wouthuysen (FW) transformation has been proposed in Ref.
\cite{FW}. Its main goals are \emph{i}) transformation of Dirac
Hamiltonians to the block-diagonal (diagonal in two spinors) form
and \emph{ii}) establishment of connection between the
relativistic quantum mechanics and the classical physics. We
utilize the method of the FW transformation for relativistic
particles in strong external fields developed in Ref. \cite{PRA}.
This method uses the expansion of the FW Hamiltonian into a power
series in the Planck constant which characterizes the order of
magnitude of quantum corrections. We present the FW Hamiltonians,
quantum mechanical and semiclassical equations of motion of
scalar, spin-1/2, and spin-1 particles and their spins in
electromagnetic fields and similar equations for spin-1/2
particles in gravitational fields and noninertial frames.

We also discuss the extension of the correspondence principle
resulting from the presented semiclassical equations.

The system of units $\hbar=c=1$ is used.

\section{Methods of Foldy-Wouthuysen Transformation}

The Foldy-Wouthuysen (FW) representation occupies a special place
in relativistic quantum mechanics thank to its unique properties.
This representation provides the best possibility of obtaining a
meaningful classical limit of the relativistic quantum mechanics
(see Refs. \cite{JMP,PRA} and references therein).

The advantages of the FW transformation can be formulated as
follows. Relations between the operators in the FW representantion
are similar to those between the respective classical quantities.
For relativistic particles in external fields, operators have the
same form as in the nonrelativistic quantum theory. For example,
the position operator is $\bm r$ and the momentum one is $\bm
p=-i\hbar\nabla$. The transition to the semiclassical description
is very simple and consists in trivial replacing operators by
corresponding classical quantities. For relativistic particles,
the connection between the square of the wave function and the
probability of a definite position of a particle is restored:
$w(\bm r)=|\Psi(\bm r)|^2$.

Initial Hamiltonian for spin-1/2 particles is
\begin{equation} {\cal H}=\beta m+{\cal E}+{\cal O},~~~
\beta{\cal E}={\cal E}\beta, ~~~\beta{\cal O}=-{\cal O}\beta.
\label{eq3} \end{equation}

Original method by Foldy and Wouthuysen \cite{FW} transforms it to
the form
\begin{equation} {\cal H}_{FW}=\beta\left( m+\frac{{\cal O}^2}{2m}-\frac{{\cal O}^4}{8m^3}
\right)+{\cal E}-\frac{1}{8m^2}\left[{\cal O},[{\cal O},{\cal
E}]\right]-\frac{i}{8m^2}\left[{\cal O},\dot{\cal O}\right].
\label{eqFW}
\end{equation}
This is a nonrelativistic transformation with relativistic
corrections. The FW transformation for relativistic spin-1/2
particles results in \cite{JMP}
\begin{equation} \begin{array}{c} {\cal H}_{FW}=\beta\epsilon+{\cal
   E}'+\frac{\beta}{4}\left\{{\cal
   O}'^2,\frac{1}{\epsilon}\right\},~~~\epsilon=\sqrt{m^2+{\cal
   O}^2},\\{\cal E}'={\cal E}-\frac14\left[\frac{\epsilon+m}
{\sqrt{2\epsilon(\epsilon+m)}},\left[\frac{\epsilon+m}
{\sqrt{2\epsilon(\epsilon+m)}},\left({\cal
E}-i\frac{\partial}{\partial t}
\right)\right]\right] \\
-\frac14\left[\frac{{\cal O}}
{\sqrt{2\epsilon(\epsilon+m)}},\left[\frac{{\cal O}}
{\sqrt{2\epsilon(\epsilon+m)}},\left({\cal
E}-i\frac{\partial}{\partial t} \right)\right]\right],\\
{\cal O}'=\frac{\beta{\cal O}}{\sqrt{2\epsilon(\epsilon+m)}}
\left({\cal E}-i\frac{\partial}{\partial t}
\right)\frac{\epsilon+m}{\sqrt{2\epsilon(\epsilon+m)}}-
\frac{\epsilon+m}{\sqrt{2\epsilon(\epsilon+m)}}\left({\cal
E}-i\frac{\partial} {\partial t}\right)\frac{\beta{\cal
O}}{\sqrt{2\epsilon(\epsilon+m)}}.
\end{array} \label{eq31} \end{equation}

The FW transformation for relativistic particles with arbitrary
spin in strong external fields has been investigated in Ref.
\cite{PRA}. The FW Hamiltonian can be expanded into the power
series in the Planck constant, $\hbar$. Initial equation is
\begin{equation} {\cal H}=\beta {\cal M}+{\cal E}+{\cal O},~~~\beta{\cal M}={\cal M}\beta,
~~~\beta{\cal E}={\cal E}\beta, ~~~\beta{\cal O}=-{\cal O}\beta.
\label{eqPRA} \end{equation} The operators ${\cal M}$ and ${\cal
E}$ are even while the operator ${\cal O}$ is odd.

In Ref. \cite{PRA}, the method developed in Ref. \cite{JMP} has
been generalized in order to take into account a possible
non-commutativity of the operators ${\cal M}$ and ${\cal O}$.
First transformation is performed with the transformation operator
$U$ defined by
\begin{equation} U=\frac{\beta\epsilon+\beta {\cal M}-{\cal
O}}{\sqrt{(\beta\epsilon+\beta {\cal M}-{\cal O})^2}}\,\beta,~~~
U^{-1}=\beta\,\frac{\beta\epsilon+\beta{\cal M}-{\cal
O}}{\sqrt{(\beta\epsilon+\beta{\cal M}-{\cal O})^2}}, ~~~
\epsilon=\sqrt{{\cal M}^2+{\cal O}^2}, \label{eq18N}
\end{equation} where $U^{-1}=U^\dagger$ when ${\cal H}={\cal
H}^\dagger$ and $U^{-1}=U^\ddagger$ when ${\cal H}={\cal
H}^\ddagger$. The latter case takes place for spinless and spin-1
particles. The sign ``$\ddagger$'' denotes the pseudo-Hermitian
conjugate and means ${\cal H}^\ddagger\equiv\beta{\cal
H}^\dagger\beta$. For particles with any spin,
$\beta\equiv\sigma_3\otimes I$, where $\sigma_3$ is the $2\times2$
Pauli matrix and $I$ is the corresponding unit matrix. The used
form of the transformation operator allows to perform the FW
transformation in the general case.

We consider the general case when external fields are
nonstationary. The exact formula for the transformed Hamiltonian
has the form
\begin{equation} \begin{array}{c} {\cal H}'=\beta\epsilon+{\cal
E}+ \frac{1}{2T}\Biggl(\left[T,\left[T,(\beta\epsilon+{\cal
F})\right]\right] +\beta\left[{\cal O},[{\cal O},{\cal
M}]\right]\\- \left[{\cal O},\left[{\cal O},{\cal F}\right]\right]
- \left[(\epsilon+{\cal M}),\left[(\epsilon+{\cal M}),{\cal
F}\right]\right] - \left[(\epsilon+{\cal M}),\left[{\cal M},{\cal
O}\right]\right]\\-\beta \left\{{\cal O},\left[(\epsilon+{\cal
M}),{\cal F}\right]\right\}+\beta \left\{(\epsilon+{\cal
M}),\left[{\cal O},{\cal F}\right]\right\} \Biggr)\frac{1}{T},
\end{array} \label{eq28N} \end{equation} where
${\cal F}={\cal E}-i\hbar\frac{\partial}{\partial t}$ and
$T=\sqrt{(\beta\epsilon+\beta{\cal M}-{\cal O})^2}$.

Hamiltonian (\ref{eq28N}) still contains odd terms proportional to
the first and higher powers of the Planck constant. This
Hamiltonian can be presented in the form \begin{equation} {\cal
H}'=\beta\epsilon+{\cal E}'+{\cal O}',~~~\beta{\cal E}'={\cal
E}'\beta, ~~~\beta{\cal O}'=-{\cal O}'\beta,
\label{eq27}\end{equation} where $\epsilon=\sqrt{{\cal M}^2+{\cal
O}^2}.$ The even and odd parts of Hamiltonian (\ref{eq27}) are
defined by the well-known relations: $${\cal
E}'=\frac12\left({\cal H}'+\beta{\cal
H}'\beta\right)-\beta\epsilon,~~~ {\cal O}'=\frac12\left({\cal
H}'-\beta{\cal H}'\beta\right).$$

Additional transformations performed according to Refs.
\cite{JMP,PRA} bring ${\cal H}'$ to the block-diagonal form. The
approximate formula for the final FW Hamiltonian is \cite{PRA}
\begin{equation}
{\cal H}_{FW}=\beta\epsilon+{\cal E}'+\frac14\beta\left\{{\cal
O}'^2,\frac{1}{\epsilon}\right\}. \label{eqf} \end{equation}

Eqs. (\ref{eq28N}),(\ref{eqf}) solve the problem of the FW
transformation for relativistic particles of arbitrary spin in
strong external fields.

Eq. (\ref{eq28N}) can be significantly simplified in some special
cases. When $[{\cal M},{\cal O}]=0$ and the external fields are
stationary, it is reduced to
\begin{equation} \begin{array}{c} {\cal H}'=\beta\epsilon+{\cal
E}+ \frac{1}{2T}\Biggl(\left[T,\left[T,{\cal E}\right]\right] \\-
\left[{\cal O},\left[{\cal O},{\cal E}\right]\right] -
\left[(\epsilon+{\cal M}),\left[(\epsilon+{\cal M}),{\cal
E}\right]\right] \\-\beta \left\{{\cal O},\left[(\epsilon+{\cal
M}),{\cal E}\right]\right\}+\beta \left\{(\epsilon+{\cal
M}),\left[{\cal O},{\cal E}\right]\right\} \Biggr)\frac{1}{T}.
\end{array} \label{eqrd} \end{equation} In this case,
$[\epsilon,{\cal M}]=[\epsilon,{\cal O}]=0$ and the operator
$T=\sqrt{2\epsilon(\epsilon+{\cal M})}$ is even.

The FW transformations becomes exact, when $[{\cal M},{\cal
O}]=0,~[{\cal O},{\cal E}]=0,$ and the external fields are
stationary. In this case \cite{PRA}
\begin{equation} {\cal H}_{FW}=\beta \epsilon+{\cal
E}, ~~~ \epsilon=\sqrt{{\cal M}^2+{\cal O}^2}. \label{eq17}
\end{equation}

The exact FW transformation can be performed in the general case
by the Eriksen method \cite{E}. The validity of the Eriksen
transformation has been argued by de Vries and Jonker \cite{VJ}.
The Eriksen transformation operator has the form \cite{E}
\begin{equation}
U=\frac12(1+\beta\lambda)\left[1+\frac14(\beta\lambda+\lambda\beta-2)\right]^{-1/2},
~~~ \lambda=\frac{{\cal H}}{({\cal H}^2)^{1/2}}, \label{E}
\end{equation} where ${\cal H}$ is the Hamiltonian in the Dirac
representation. This operator brings the Dirac wave function and
the Dirac Hamiltonian to the FW representation in one step.
However, it is difficult to use the Eriksen method for obtaining
an explicit form of the relativistic FW Hamiltonian because the
general final formula is very cumbersome and contains roots of
Dirac matrix operators. Therefore, the Eriksen method was not used
for relativistic particles in external fields.

Other methods of the FW transformation have been developed in
Refs. \cite{Pa,B,Neznamov,Gos}.

\section{Quantum Mechanical and Semiclassical Equations of Motion of
Particles and their Spins}

\subsection{Equations of Motion in Electromagnetic Fields}

FW Hamiltonians have been derived for relativistic scalar
particles \cite{PRA,TMP2008}, relativistic spin-1/2 particles with
electric and magnetic dipole moments \cite{PRA} in strong
electromagnetic fields, and relativistic spin-1 particles without
electric dipole moments (EDMs) \cite{EPJC} in a strong uniform
magnetic field.

In the FW representation, the transition to the semiclassical
approximation becomes trivial. It consists in replacing operators
by corresponding classical quantities. The quantum mechanical
equations of motion of particles and their spins are given by
\begin{equation}\frac{d\bm\pi}{dt}=\frac{i}{\hbar}[{\cal
H}_{FW},\bm\pi] -\frac{e}{c}\cdot\frac{\partial\bm A}{\partial t},
~~~ \bm\pi=\bm p-\frac{e}{c}\bm A, \label{eqme} \end{equation}
\begin{equation}\frac{d\bm\Pi}{dt}=\frac{i}{\hbar}[{\cal
H}_{FW},\bm\Pi], \label{eqpoe} \end{equation} where $\bm\pi$ is
the kinetic momentum operator and $\bm\Pi$ is the polarization
operator. Usual definitions of the Dirac matrices are applied.

The equation of spin-1/2 particle motion in the strong
electromagnetic field to within first-order terms in the Planck
constant has the form \cite{PRA} \begin{equation}
\begin{array}{c} \frac{d\bm \pi}{dt}=e\bm
E+\beta\frac{ec}{4}\left\{\frac{1}{\epsilon'},
\left([\bm\pi\times\bm B]-[\bm B\times\bm\pi]\right)\right\}
%\\
+\mu' \nabla(\bm\Pi\cdot\bm B)+
\frac{\mu_0}{2}\left\{\frac{mc^2}{\epsilon'},
\nabla(\bm\Pi\cdot\bm
H)\right\}\\-\frac{\mu'c}{4}\left\{\frac{1}{\epsilon'},
\left[\nabla(\bm\Sigma\cdot[\bm\pi\times\bm E])-
\nabla(\bm\Sigma\cdot[\bm E\times\bm\pi]) \right]\right\}\\-
\frac{\mu_0mc^3}{\sqrt{2\epsilon'(\epsilon'+mc^2)}}\left[\nabla(\bm\Sigma\cdot[\bm\pi\times\bm
E])- \nabla(\bm\Sigma\cdot[\bm
E\times\bm\pi])\right]\frac{1}{\sqrt{2\epsilon'(\epsilon'+mc^2)}}\\
-\frac{\mu'c^2}{2\sqrt{2\epsilon'(\epsilon'+mc^2)}}\left\{(\bm{\Pi}\cdot\bm\pi),\left[\nabla
(\bm{H}\cdot\bm\pi)+\nabla(\bm{\pi}\cdot\bm B)\right]\right\}
\frac{1}{\sqrt{2\epsilon'(\epsilon'+mc^2)}}. \end{array}
\label{eq35} \end{equation}

This equation can be divided into two parts. The first part does
not contain the Planck constant and describes the quantum
equivalent of the Lorentz force. The second part is of order of
$\hbar$. This part defines the relativistic expression for the
Stern-Gerlach force. Small terms proportional to $d$ are omitted.

The equation of spin motion is given by \cite{PRA}
\begin{equation} \begin{array}{c} \frac{d\bm{\Pi}}{dt}=\frac{2\mu'}{\hbar}\bm\Sigma\times\bm B+\frac{\mu_0}{\hbar} \left\{\frac{mc^2}{\epsilon'},
\bm\Sigma\times\bm B\right\} %\\
-\frac{\mu'c}{2\hbar}\left\{\frac{1}{\epsilon'},
\left[\bm\Pi\times(\bm \pi\times \bm E)-\bm\Pi\times(\bm E\times \bm\pi) \right]\right\}\\
-
\frac{\mu_0mc^3}{\hbar\sqrt{\epsilon'(\epsilon'+mc^2)}}\left[\bm\Pi\times(\bm
\pi\times \bm E)
-\bm\Pi\times(\bm E\times \bm\pi) \right]\frac{1}{\sqrt{\epsilon'(\epsilon'+mc^2)}}\\
- \frac{\mu'c^2}{\hbar\sqrt{2\epsilon'(\epsilon'+mc^2)}}
\left\{(\bm\Sigma\times\bm \pi),
(\bm{H}\cdot\bm\pi+\bm{\pi}\cdot\bm B)\right\}
\frac{1}{\sqrt{2\epsilon'(\epsilon'+mc^2)}} \\
+\frac{2d}{\hbar}\bm\Sigma\times\bm E %\\
-\frac{dc^2}{\hbar\sqrt{2\epsilon'(\epsilon'+mc^2)}}
\left\{(\bm\Sigma\times\bm \pi),
(\bm{E}\cdot\bm\pi+\bm{\pi}\cdot\bm E)\right\}
\frac{1}{\sqrt{2\epsilon'(\epsilon'+mc^2)}}
\\+\frac{dc}{2\hbar}\left\{\frac{1}{\epsilon'}, \left[\bm\Pi\times(\bm
\pi\times \bm B)-\bm\Pi\times(\bm B\times \bm\pi) \right]\right\}.
\end{array} \label{eq36} \end{equation}

%%%%%%%%%%%%%%%%%%%%%%%%%!!!!!!!!!!!!!!!!!!!!!!!!!!!!!!!!!!!!!!!!!!!
Eqs. (\ref{eq35}),(\ref{eq36}) describe strong-field effects.

For spinless particles, the operator equation of particle motion
takes the form \cite{PRA}
\begin{equation} \begin{array}{c} \frac{d\bm \pi}{dt}=e\bm
E+\beta\frac{ec}{4}\left\{\frac{1}{\epsilon},
\left([\bm\pi\times\bm B]-[\bm B\times\bm\pi]\right)\right\}.
\end{array} \label{eqpz} \end{equation}

The right hand side of this equation coincides with the
spin-independent part of the corresponding equation for spin-1/2
particles. Additional terms in the operator equation of particle
motion derived in Ref. \cite{TMP2008} are of order of $\hbar^2$.

Similar equations have been derived for spin-1 particles in a
strong magnetic field. When the matrices $\bm\Sigma=I\bm S$ ($\bm
S$ is the $3\times3$ spin matrix) and $\bm\Pi=\rho_3\bm S$ acting
on the functions $\phi$ and $\chi$ are entered, the operator
equations of motion of spin-1 particles and their spins are given
by \cite{EPJC}
\begin{equation} \begin{array}{c} \frac{d\bm
\pi}{dt}=-\rho_3\frac{ec}{2}\left\{\!\frac{1}{\epsilon'}, \bm
B\times\bm\pi\!\right\}
%\\
-\frac{e^2\hbar c^2}{2}(\bm\Pi\cdot\bm
B)\left\{\!\frac{1}{{\epsilon'}^3}, \bm B\times\bm\pi\!\right\}\\
+\frac{e^2\hbar(g-2)}{2m}(\bm
B\cdot\bm\pi)\left[\frac{\bm\Pi\times\bm
B}{\epsilon'(\epsilon'+mc^2)} \right.\\\left.
+\frac{c^2}{4}\left\{\frac{2\epsilon'+mc^2}{{\epsilon'}^3(\epsilon'+mc^2)^2},\left\{(\bm
B\times\bm\pi), (\bm\Pi\cdot\bm\pi)\right\} \right\}\right],
\end{array} \label{eqVM} \end{equation}
\begin{equation} \begin{array}{c} \frac{d\bm{\Pi}}{dt}=\left[
\frac{e(g-2)}{2mc}+\frac{ec}{\epsilon'}\right] \bm\Sigma\times\bm
B \\
-\frac{ec(g-2)}{4m}(\bm{B}\cdot\bm\pi)\left\{\frac{1}{\epsilon'(\epsilon'+mc^2)},
\bm\Sigma\times\bm \pi\right\}.
\end{array} \label{eqVP} \end{equation}

Eqs. (\ref{eqVM}) and (\ref{eqVP}) are derived with allowance for
terms up to the first and zero orders in the Planck constant,
respectively. As a result, Eq. (\ref{eqVM}) describes the quantum
equivalent of the Lorentz force and defines the relativistic
expression for the Stern-Gerlach force. The Stern-Gerlach force is
rather weak in the uniform magnetic field because it is of order
of $B^2$.

The semiclassical limit of the relativistic quantum mechanics has
been investigated in Ref. \cite{PRA}. The expansion into a power
series in the Planck constant can be available only if
\begin{equation} pl\gg\hbar, \label{rel1} \end{equation} where $p$
is the momentum of the particle and $l$ is the characteristic size
of the nonuniformity region of the external field. This relation
is equivalent to \begin{equation} \lambda\ll l, \label{rel2}
\end{equation} where $\lambda$ is the de Broglie wavelength. Eqs.
(\ref{rel1}),(\ref{rel2}) result from the fact that the Planck
constant appears in the final Hamiltonian due to commutators
between the operators ${\cal M},{\cal E}$, and ${\cal O}$.

One needs to average the operators in the quantum mechanical
equations. When the FW representation is used and relations
(\ref{rel1}),(\ref{rel2}) are valid, the semiclassical transition
consists in trivial replacing operators with corresponding
classical quantities. If the momentum and position operators are
chosen to be the dynamical variables, relations
(\ref{rel1}),(\ref{rel2}) are equivalent to the condition
\begin{equation} |<p_i>|\cdot|<x_i>|\gg|<[p_i,x_i]>|=\hbar,~~~
i=1,2,3. \label{rel3} \end{equation} The angular brackets which
designate averaging in time will be hereinafter omitted.

%Obtained semiclassical equations may differ from corresponding
%classical ones.

As a result of replacing operators by corresponding classical
quantities, the semiclassical equations of motion of spin-1/2
particles and their spins take the form \begin{equation}
\begin{array}{c} \frac{d\bm \pi}{dt}=e\bm E+\frac{ec}{\epsilon'}
\left(\bm\pi\times\bm H\right)
%\\
+\mu'\nabla(\bm P\cdot\bm H)+ \frac{\mu_0}{mc^2\epsilon'} \nabla(\bm P\cdot\bm H)\\
-\frac{\mu'c}{\epsilon'} \nabla(\bm P\cdot[\bm\pi\times\bm E])%\\
-\frac{\mu_0mc^3}{\epsilon'(\epsilon'+mc^2)}\nabla(\bm
P\cdot[\bm\pi\times\bm E])\\
-\frac{\mu'c^2}{\epsilon'(\epsilon'+mc^2)}(\bm{P}\cdot\bm\pi)\nabla
(\bm{H}\cdot\bm\pi), ~~~~~~~ \bm P=\frac{\bm s}{S}, \end{array}
\label{eqw} \end{equation}
\begin{equation} \begin{array}{c} \frac{d\bm P}{dt}=2\mu'\bm P\times\bm H+ \frac{2\mu_0mc^2}{\epsilon'}( \bm P\times\bm H)%\\
-\frac{2\mu'c}{\epsilon'} \left(\bm P\times[\bm \pi\times \bm
E]\right)\\ -\frac{2\mu_0mc^3}{\epsilon'(\epsilon'+mc^2)}\left(\bm
P\times[\bm \pi\times \bm E] \right)
%\\
-\frac{2\mu'c^2}{\epsilon'(\epsilon'+mc^2)} (\bm P\times\bm \pi)(\bm{\pi}\cdot\bm H)\\ %
+2d\bm P\times\bm E-\frac{2dc^2}{\epsilon'(\epsilon'+mc^2)} (\bm
P\times\bm \pi)(\bm{\pi}\cdot\bm E) %\\
+\frac{2dc}{\epsilon'} \left(\bm P\times[\bm \pi\times \bm
H]\right). \end{array} \label{eqt} \end{equation} In Eqs.
(\ref{eqw}),(\ref{eqt}), $\epsilon'=\sqrt{m^2c^4+c^2\bm{\pi}^2}$,
$\bm P$ is the polarization vector, $\bm s$ is the spin vector
(i.e., the average spin), and $S$ is the spin quantum number.

Similar semiclassical equation of spin motion for spin-1 particles
has been derived in Ref. \cite{EPJC}.

For scalar particles
\begin{equation} \begin{array}{c}
\frac{d\bm \pi}{dt}=e\bm E+\frac{ec}{\sqrt{m^2c^4+c^2\bm\pi^2}}
\left(\bm\pi\times\bm H\right). \end{array} \label{eqwl}
\end{equation}

Two first terms in right hand sides of Eqs.
(\ref{eqw}),(\ref{eqwl}) are the same as in the classical
expression for the Lorentz force. This is a manifestation of the
Niels Bohr's correspondence principle. The part of Eq. (\ref{eqt})
dependent on the magnetic moment coincides with the well-known
Thomas-Barg\-mann-Michel-Te\-leg\-di (T-BMT) equation. It is
natural because the T-BMT equation has been derived without the
assumption that the external fields are weak. The whole Eq.
(\ref{eqt}) coincides with the corresponding classical equation
derived in Ref. \cite{EDMNSPC}. The relativistic formula for the
Stern-Gerlach force can be obtained from the Lagrangian consistent
with the T-BMT equation (see Ref. \cite{PK}). The semiclassical
and classical formulas describing this force also coincide.
High-order corrections in $\hbar$ to the quantum equations of
motion of particles and their spins bring a difference between
quantum and classical approaches.

\subsection{Equations of Motion in Gravitational Fields and Noninertial Frames}

The best compliance between the description of spin effects in the
classical and quantum gravity has been proved in Refs.
\cite{PRD,PRD2}. In these works, some Hamiltonians in the Dirac
representation derived in Refs. \cite{Ob1,HN} from the initial
covariant Dirac equation have been used. The initial Dirac
Hamiltonians have been transformed to the FW representation by the
method elaborated in Ref. \cite{JMP}.

The exact transformation of the Dirac equation for the metric
\begin{equation}
ds^2=V^2(\bm r)(dx^0)^2-W^2(\bm r)(d\bm r\cdot\bm r)
\label{metric}\end{equation} to the Hamiltonian form has been
carried out by Obukhov \cite{Ob1}:
\begin{equation}
i\frac {\partial \psi}{\partial t}={\cal H}\psi, ~~~ {\cal H} =
\beta m V+\frac12\{{\cal F},\bm\alpha\cdot\bm p\},
\label{eqO}\end{equation} where ${\cal F} = V/W$. Hamiltonian
(\ref{eqO}) covers several cases including a weak Schwarzschild
field in the isotropic coordinates and a uniformly accelerated
frame.

The relativistic FW Hamiltonian derived in Ref. \cite{PRD} has the
form
\begin{equation}
\begin{array}{c}
{\cal H}_{FW}=\beta\epsilon
+\frac{\beta}{2}\left\{\frac{m^2}{\epsilon },V-1\right\}
+\frac{\beta}{2}\left\{\frac{\bm
p^2}{\epsilon },{\cal F}-1\right\}\\
-\frac{\beta m}{4 \epsilon (\epsilon
+m)}\biggl[\bm{\Sigma}\cdot(\bm\phi\times\bm p)-
\bm{\Sigma}\cdot(\bm p\times\bm\phi)+ \nabla\!
\cdot\!\bm\phi\biggr]
\nonumber\\
+\frac{\beta m(2\epsilon ^3+2\epsilon ^2m+2\epsilon
m^2+m^3)}{8\epsilon ^5 (\epsilon +m)^2}(\bm p\cdot\!\nabla)(\bm
p\cdot\!\bm\phi)\\+ \frac{\beta}{4\epsilon
}\left[\bm{\Sigma}\cdot(\bm f\times\bm p)- \bm{\Sigma}\cdot(\bm
p\times\bm f)+\nabla\! \cdot\!\bm f\right]-\frac{\beta(\epsilon
^2+m^2)}{4\epsilon ^5}(\bm p\cdot\!\nabla)(\bm p\cdot\!\bm f),
\label{eq7}\end{array}\end{equation} where $\epsilon=\sqrt{m^2+\bm
p^2},~\bm\phi=\nabla V,~\bm f=\nabla{\cal F}$.

The operator equations of momentum and spin motion take the form
\cite{PRD}
\begin{eqnarray}
\frac{d\bm p}{dt}=i[{\cal H}_{FW},\bm p]=
-\frac{\beta}{2}\left\{\frac{m^2}{\epsilon
},\bm\phi\right\}-\frac{\beta}{2}\left\{\frac{\bm p^2}{\epsilon
},\bm f\right\}
\nonumber\\
+\frac{m}{2 \epsilon (\epsilon
+m)}\nabla\bigl(\bm{\Pi}\cdot(\bm\phi\times\bm p)\bigr) -
\frac{1}{2\epsilon } \nabla\bigl(\bm{\Pi}\cdot(\bm f\times\bm
p)\bigr) \label{eq11}\end{eqnarray} and
\begin{eqnarray}
\frac{d\bm\Pi}{dt}=\frac{m}{\epsilon (\epsilon
+m)}\bm\Sigma\times\left(\bm\phi\times\bm p\right)
%\nonumber\\
-\frac{1}{\epsilon }\bm\Sigma\times\left(\bm f\times\bm p\right),
\label{eq12}\end{eqnarray} respectively.

The semiclassical equations of motion are \cite{PRD}
\begin{eqnarray}
\frac{d\bm p}{dt}= -\frac{m^2}{\epsilon }\bm\phi-\frac{\bm
p^2}{\epsilon }\bm f+\frac{m}{2 \epsilon (\epsilon
+m)}\nabla\bigl(\bm{P}\cdot(\bm\phi\times\bm p)\bigr)
%\nonumber\\
- \frac{1}{2\epsilon } \nabla\bigl(\bm{P}\cdot(\bm f\times\bm
p)\bigr) \label{eq14}\end{eqnarray} and
\begin{equation}
\frac{d\bm P}{dt}= \frac{m}{\epsilon (\epsilon +m)}\bm
P\times\left(\bm\phi\times\bm p\right)- \frac{1}{\epsilon }\bm
P\times\left(\bm f\times\bm p\right), \label{eq15}\end{equation}
respectively. In Eq. (\ref{eq14}), two latter terms describe the
force dependent on the spin. This force is similar to the
electromagnetic Stern-Gerlach force and is rather weak. The
angular velocity of spin rotation is given by
\begin{equation}
\bm\Omega=-\frac{m}{\epsilon (\epsilon +m)}\left(\bm\phi\times\bm
p\right)+\frac{1}{\epsilon}\left(\bm f\times\bm p\right).
\label{eqom}\end{equation}

We can find similar equations describing a change of the direction
of particle momentum, $\bm n=\bm p/p\;$:
\begin{equation}
\frac{d\bm n}{dt}=\bm\omega\times\bm n, ~~~ \bm\omega=
\frac{m^2}{\epsilon p}\bigl( \bm\phi\times\bm
n\bigr)+\frac{p}{\epsilon}\bigl(\bm f\times\bm n\bigr).
\label{eq18}\end{equation}

Explicit form of the equation of motion of the three-component
spin has been obtained by Pomeransky and Khriplovich \cite{PK}.
The derivation of this equation is based on neglecting the
relatively weak influence of the spin on a particle's trajectory
which results in a weak violation of the equivalence principle by
the curvature-dependent terms \cite{Plyatsko}. In this
approximation, the Pomeransky-Khriplovich equations (PKEs) for the
three-component and four-component spins agree with the seminal
Mathisson-Papapetrou equations for the four-component spin
\cite{MathissonPapapetrou} (see Ref. \cite{Warszawa} and
references therein).

A simple calculation shows \cite{PRD} that Eqs.
(\ref{eq14})--(\ref{eq18}) coincide with the corresponding
classical equations of motion of particles and their spins
obtained from the PKEs for given metric (\ref{metric}). The
gravitational analogue of the Stern-Gerlach force defined by Eq.
(\ref{eq14}) coincides with the corresponding force obtained from
the PKEs for the three-component spin (see Ref. \cite{Warszawa}).

The FW Hamiltonian and the operators of velocity and acceleration
have also been calculated for the Dirac particle in the rotating
frame \cite{PRD2}. The exact Dirac Hamiltonian derived in Ref.
\cite{HN} has been used. In Ref. \cite{PRD2}, perfect agreement
between classical and quantum approaches has also been
established. The operators of velocity and acceleration are equal
to
\begin{eqnarray}
\bm v=\beta\frac{\bm p}{\epsilon}-\bm\omega\times\bm r, ~~~
\epsilon =\sqrt{m^2+\bm p^2},
\nonumber\\
\bm w=2\beta\frac{\bm p\times\bm\omega}{\epsilon}
+\bm\omega\times(\bm\omega\times\bm r)
%\nonumber\\
=2\bm v\times\bm\omega-\bm\omega\times(\bm\omega\times\bm r).
\label{eqvn}\end{eqnarray} Quantum mechanical formula (\ref{eqvn})
for the acceleration of the relativistic spin-1/2 particle
coincides with the classical formula \cite{Gol} for the sum of the
Coriolis and centrifugal accelerations. Obtained results also
agree with the corresponding nonrelativistic formulas from
\cite{HN}.

Quantum equations of motion of Dirac particles and their spins in
a gravitational field of a rotating body defined by the
Lense-Thirring metric being a weak field limit of the Kerr metric
have been derived in Ref. \cite{PRD3}. The equation of rotation of
the spin contains two parts. One of them is defined by the static
part of the Lense-Thirring (LT) metric and is expressed by Eqs.
(\ref{eq12}) and (\ref{eq15}). The second part is proportional to
the total angular momentum of the source, $\bm J=Mca\bm e_z$, and
is defined by the operator of angular velocity of the spin
precession \cite{PRD3}:
\begin{eqnarray}
\bm\Omega_{LT} &=& \frac{G}{c^2r^3} \left[\frac{3(\bm r\cdot\bm
J)\bm r} {r^2} - \bm J\right]
-\frac{3G}{4}\left\{\frac{1}{\epsilon (\epsilon +
mc^2)},\left[\frac{2\{\bm l,(\bm J\cdot\bm l)\}}{r^5}
\right.\right.\nonumber\\
&& \left.\left. \,+\,\frac{1}{2}\left\{(\bm p\times\bm l - \bm l
\times\bm p), \frac{(\bm r\cdot\bm J)}{r^5}\right\} + \left\{(\bm
p \times(\bm p \times\bm J)),\frac{1}{r^3}\right\}\right]\right\}.
\label{finalOmega}
\end{eqnarray}

The semiclassical formula corresponding to Eq. (\ref{finalOmega})
and describing the motion of average spin has the form \cite{PRD3}
\begin{equation}\label{OmegaVt}
\bm\Omega_{LT}= \frac{G}{c^2r^3} \left[\frac{3(\bm r\cdot\bm J)\bm
r} {r^2} - \bm J\right] -
\frac{3G}{r^3\epsilon(\epsilon+mc^2)}\left[ \frac{2\bm l (\bm
J\cdot\bm l)+(\bm p\times\bm l) (\bm r\cdot\bm J)} {r^2}+\bm
p\times(\bm p \times\bm J)\right].
\end{equation}
This equation can also be expressed in the equivalent form
\cite{PRD3}:
\begin{equation}\label{OmegaCl}
\bm\Omega^{(2)}=\frac{G}{c^2r^3}\left[\frac{3(\bm r\cdot\bm J)\bm
r}{r^2} - \bm J\right] -
\frac{3G}{r^5\epsilon(\epsilon+mc^2)}\left[\bm l (\bm l \cdot\bm
J)+(\bm r\cdot\bm p) (\bm p\times(\bm r\times\bm J))\right].
\end{equation}

The equation of motion of the particle defines the evolution of
the contravariant four-momentum operator which spatial components
($a,b=1,2,3$) are given by
$$p^a = g^{ab}p_b + g^{0a}p_0.$$

In a stationary metric, the evolution of the contravariant
momentum operator in the weak field approximation is defined by
\begin{eqnarray}
F^a=\frac{dp^a}{dt}=-\frac{dp_a}{dt}+\frac14\left\{\left\{v^b,
\frac{\partial g^{ai}}{\partial x^b}\right\},p_i\right\},\qquad
\frac{d\bm p}{dt}=\frac{i}{\hbar}[{\cal H}_{FW},\bm p],
\label{eqFi}\end{eqnarray} where $F^a$ is the force operator and
$v^a\approx\beta c^2p^a/\epsilon \approx c^2p^a/{\cal H}_{FW}$ is
the velocity operator.

The force operator caused by the LT effect is equal to \cite{PRD3}
\begin{eqnarray}
\bm F=\frac c2 \left({\rm curl}\,\bm K\times\bm p-\bm p\times {\rm
curl}\,\bm K\right)+\bm F_s,\label{finalpi}
\end{eqnarray}
where
\begin{eqnarray}
{\rm curl}\,\bm K=\frac{2G}{c^3r^3} \left[\frac{3(\bm r\cdot
\bm J)\bm r} {r^2} - \bm J\right],\quad %\nonumber\\
\bm F_s=-\nabla\left(\frac{\hbar G}{2c^2r^3} \left[\frac{3(\bm
r\cdot\bm J)(\bm r\cdot\bm \Sigma)}{r^2}- \bm J\cdot\bm
\Sigma\right]\right.\nonumber\\-\frac{3\hbar G }{8}\left\{\frac
{1}{\epsilon(\epsilon+mc^2)},\left[\frac{2\{(\bm J\cdot\bm l),
(\bm \Sigma\cdot\bm l)\}}{r^5}+\frac{1}{2} \left\{\left(\bm\Sigma
\cdot (\bm p\times\bm l)-\bm \Sigma\cdot(\bm l\times\bm p)\right),
\frac {(\bm r\cdot\bm J)}{r^5}\right\}\right.\right.\nonumber\\
\left.\left.\left. +\left\{\bm \Sigma\cdot(\bm p\times(\bm
p\times\bm J)), \frac{1}{r^3}\right\}\right]\right\}\right).
\label{finlK}
\end{eqnarray} These equations are given without allowance for
contributions from $V,W$. The part of operator equations
(\ref{finalpi}) and (\ref{finlK}) defining the spin-independent
force is in the best compliance with the corresponding classical
equation \cite{LL}. Since the Dirac spin operator is $\bm
s=\hbar\bm \Sigma/2$, Eqs. (\ref{finalpi}) and (\ref{finlK}) yield
the corresponding semiclassical equation \cite{PRD3}:
\begin{eqnarray}
\bm{\mathcal{F}}=c\,{\rm curl}\,\bm K\times\bm
p+\bm{\mathcal{F}}_s,\label{finalpt}
\end{eqnarray}
\begin{eqnarray}
\bm{\mathcal{F}}_s=-\nabla\left(\frac{G}{c^2r^3} \left[\frac{3(\bm
r\cdot\bm J)(\bm r\cdot\bm s)}{r^2}- \bm J\cdot\bm s\right]\right.\nonumber\\
\left.-\frac{3G}{\epsilon(\epsilon+mc^2)}\left[\frac{2(\bm
J\cdot\bm l)(\bm s\cdot\bm l)}{r^5}+\frac{\left(\bm s \cdot [\bm
p\times\bm l]\right) (\bm r\cdot\bm J)}{r^5} + \frac{\left(\bm s
\cdot [\bm p\times[\bm p\times\bm J]]\right)}{r^3}\right]\right).
\label{finlt}
\end{eqnarray}

The relativistic result (\ref{finlK}), (\ref{finlt}) for the
spin-dependent force perfectly agrees with the corresponding
nonrelativistic classical formulas previously obtained in Ref.
\cite{Wald} on the basis of the Mathisson-Papapetrou equations
\cite{MathissonPapapetrou}.

The presented quantum equations agree with the classical results
obtained with the PKEs. This follows from the fact that the
spin-dependent part of the Hamiltonian has the form ${\cal
H}_s=\hbar(\bm\Omega^{(1)} \cdot\bm
\Sigma+\bm\Omega^{(2)}\cdot\bm\Pi)/2$ that perfectly agrees with
the general classical Eq. (47) of Ref. \cite{PK}.

Thus, the classical and quantum approaches are in full agreement.
Pu\-re\-ly quantum effects are not too important. They consist in
appearing some additional terms in the FW Hamiltonian. However,
the leading corrections are proportional to derivatives of
$\bm\phi$ and $\bm f$ and are similar to the well-known Darwin
term in the electrodynamics. As a result, the influence of the
additional terms on the motion of particles and their spins in
gravitational fields can be neglected. In this case, the classical
and semiclassical equations of motion of particles and their spins
coincide.

\section{Foldy-Wouthuysen Transformation and Niels Bohr's Correspondence Principle}

The correspondence principle has been formulated by Niels Bohr
\cite{NielsBohr}. This principle is very important because it
establishes a connection between classical and quantum physics.
The correspondence principle states that the behavior of quantum
mechanical systems reproduces the classical physics in the limit
of large quantum numbers \cite{NielsBohr}. As follows from this
statement, the quantum mechanics should generate classical results
in the limit of $S\rightarrow\infty$ only.

The classical limit of the relativistic quantum mechanics can be
obtained with averaging operators (i.e., substituting classical
quantities for corresponding operators), neglecting commutators,
and vanishing the Planck constant. For particles with any spins,
the problem can be unambiguously solved with the FW transformation
because operators in the FW representation have the same form as
in the nonrelativistic quantum theory.

When high-order terms in $\hbar$ are omitted, the semiclassical
equations of motions of particles and their spins coincide with
the corresponding classical equations. Since particles with spin
0, 1/2, and 1 are considered, one can conclude that \textbf{the
behavior of quantum mechanical systems reproduces the classical
physics in the limit of large quantum numbers for particles with
arbitrary spin.} Thus, there is no need for the additional
restriction $S\rightarrow\infty$. The Niels Bohr's correspondence
principle is valid for particles with any spin as well as for
spinless particles. This is an evident extension of the
correspondence principle because the definitions of the spin in
the initial quantum mechanical equations and classical equations
significantly differ. The spin is defined by vector sums of spin
matrices in the quantum mechanics and the intrinsic angular
momentum in the classical physics. The form of the spin matrices
depends on the spin quantum number $S$. So, the obtained extension
of the Niels Bohr's correspondence principle is rather nontrivial.

\section{Discussion and Summary}

We can conclude that the FW transformation for relativistic
particles in strong external fields gives the meaningful classical
limit of the relativistic quantum mechanics. We have established
that the semiclassical equations of motion of particles and their
spins coincide with the corresponding classical equations. The
Niels Bohr's correspondence principle is valid not only in the
limit of large spin quantum numbers but also for particles with
any spin as well as for spinless particles.

To confirm the consistency of classical and quantum equations, we
can additionally use the results obtained by Pomeransky and
Khriplovich \cite{PK}. In this work, the quantum Lagrangians of
interaction of arbitrary spin particles with electromagnetic and
gravitational fields have been derived by the method of scattering
amplitudes. The comparison of these Lagrangians with the
corresponding classical ones shows the coincidence of terms of the
zeroth and first orders in spin.

\section*{Acknowledgments}

The author is grateful to V. P. Neznamov, O. V. Teryaev and Yu. N.
Obukhov for helpful discussions and comments. This work was
supported by the Belarusian Republican Foundation for Fundamental
Research (Grant No. $\Phi$08D-001).

\end{document}